\documentclass[12pt,article]{JHEP3}
\usepackage{amsmath,amssymb,amsfonts}
\usepackage{subfigure,graphicx,url}
\usepackage{epsfig,xspace}

\def\lfig#1#2#3#4{
\begin{figure}
\centerline{\hfill \includegraphics[height=#3]{#2}\hfill}
\caption{#1 \label{#4}}
\end{figure}
}

\numberwithin{equation}{section}

\def\varpi{\zeta}

\def\Im{\,{\rm Im}\,}
\def\Re{\,{\rm Re}\,}

\def\({\left(}
\def\){\right)}
\def\[{\left[}
\def\]{\right]}
\def\<{\left\langle}
\def\>{\right\rangle}
\def\hf{{1\over 2}}

\renewcommand{\d}{\mathrm{d}}
\newcommand{\de}{\mathrm{d}}

\newcommand{\I}{\mathrm{i}}

\newcommand{\cL}{\mathcal{L}}

\def\vph{\varphi}
\newcommand{\p}{\partial}

\newcommand{\half}{\frac{1}{2}}

\newcommand{\cF}{\mathcal{F}}

\newcommand{\cD}{\mathcal{D}}
\newcommand{\cS}{\mathcal{S}}

\newcommand{\cM}{\mathcal{M}}
\newcommand{\cW}{\mathcal{W}}

\newcommand{\cX}{\mathcal{X}}
\newcommand{\CX}{\mathcal{X}}
\newcommand{\cR}{\mathcal{R}}

\newcommand{\cJ}{\mathcal{J}}
\newcommand{\cZ}{\mathcal{Z}}

\newcommand{\cO}{\mathcal{O}}

\newcommand{\cU}{\mathcal{U}}

\newcommand{\cQ}{\mathcal{Q}}

\DeclareSymbolFont{AMSa}{U}{msa}{m}{n}
\DeclareSymbolFont{AMSb}{U}{msb}{m}{n}
\DeclareMathSymbol{\fieldR}{\mathalpha}{AMSb}{"52}

\newcommand{\kahler}{{K\"ahler}\xspace}

\newcommand{\qk}{{quaternionic-K\"ahler}\xspace}

\newcommand{\nn}{\nonumber}

\newcommand{\eps}{\epsilon}
\newcommand{\IR}{\mathbb{R}}
\newcommand{\IC}{\mathbb{C}}
\newcommand{\IZ}{\mathbb{Z}}

\newcommand{\tzeta}{\tilde\zeta}

\newcommand{\txi}{\tilde\xi}

\newcommand{\CP}{\IC P^1}

\def\bea{\begin{eqnarray}}
\def\eea{\end{eqnarray}}
\def\be{\begin{equation}}
\def\ee{\end{equation}}
\def\ba{\begin{align}}
\def\ea{\end{align}}
\def\bse{\begin{subequations}}
\def\ese{\end{subequations}}

\def\bF{\bar F}
\def\bY{\bar Y}

\def\bZ{\bar Z}

\def\ba{\bar a}
\def\bb{\bar b}
\def\bc{\bar c}

\def\bi{\bar \imath}

\def\bw{\bar w}

\def\bz{\bar z}
\def\bu{\bar u}

\newcommand{\CA}{{\cal{A}}}
\newcommand{\CB}{{\cal{B}}}
\newcommand{\CC}{{\cal{C}}}
\newcommand{\CD}{{\cal{D}}}

\def\ui#1{^{[#1]}}
\def\di#1{_{[#1]}}
\def\ci#1{c^{[#1]}}
\def\cij#1{c^{[#1]}}
\def\mui#1{\mu^{[#1]}}
\def\txii#1{{\tilde\xi}^{[#1]}}
\def\ai#1{{\alpha}^{[#1]}}
\def\nui#1{\nu_{[#1]}}
\def\xii#1{\xi_{[#1]}}

\def\Hij#1{H^{[#1]}}

\def\hHij#1{\hat H^{[#1]}}

\def\Xiijg#1#2{\Xi_{\gamma_{#2}}^{[#1]}}
\def\Xigi#1{\Xi_{\gamma_{#1}}}
\def\Xigmi#1{\Xi_{-\gamma_{#1}}}

\newcommand{\Li}{{\rm Li}}

\def\Thkl{\Theta_{\gamma}}

\def\Xikl{\Xi_{\gamma}}

\def\hnkl{n_{\gamma}}

\def\Ggs{G_{\gamma}}

\def\ellg#1{\ell_{#1}}

\def\Thkli#1{\Theta_{\gamma_{#1}}}

\def\hng#1{n_{\gamma_{#1}}}

\def\Igg#1{\cJ_{\gamma_{#1}}}

\def\Ggss#1{G_{#1}}

\def\mui#1{\mu^{[#1]}}
\def\nui#1{\nu_{[#1]}}
\def\etai#1{\eta_{[#1]}}

\def\psit{\psi_\tau}
\def\tap{\tau'}
\def\btap{\bar\tau'}

\def\xip{\xi_{(\rm pert)}}
\def\txip{\tilde\xi^{(\rm pert)}}

\def\XXint#1#2#3{{\setbox0=\hbox{$#1{#2#3}{\int}$}
\vcenter{\hbox{$#2#3$}}\kern-.5\wd0}}

\def\cij#1{c}
\def\ci#1{c}

\newcommand{\hCX}{\mathcal{X}}

\def\hHij#1{H^{[#1]}}

\preprint{LPTA/10-020
}
\title{TBA for non-perturbative moduli spaces}
\author{Sergei Alexandrov and Philippe Roche
\\
{\it Laboratoire de Physique Th\'eorique \&
Astroparticules, CNRS UMR 5207, \\
Universit\'e Montpellier II, 34095 Montpellier Cedex 05, France}
}
\abstract{Recently, an exact description of instanton corrections to the moduli spaces
of 4d $N=2$ supersymmetric gauge theories
compactified on a circle and Calabi-Yau compactifications of Type II superstring theories was found.
The equations determining the instanton contributions turn out to have the form of Thermodynamic Bethe Ansatz.
We explore further this relation and, in particular,
we identify the contact potential of quaternionic string moduli space with the free energy
of the integrable system and the K\"ahler potential of the gauge theory moduli space with
the Yang-Yang functional.
We also show that the corresponding S-matrix satisfies all usual constraints of 2d integrable models,
including crossing and bootstrap, and derive the associated Y-system.
Surprisingly, in the simplest case the Y-system is described by the MacMahon function
relevant for crystal melting and topological strings.
}

\begin{document}

\section{Introduction}

The last years have marked a remarkable manifestation of integrability and in particular
of (Thermodynamic) Bethe Ansatz in description of dynamics of gauge and string theories.
A prominent example is the calculation of the spectrum of conformal dimensions in $N=4$
super Yang-Mills theory \cite{Minahan:2002ve} and the exact solution of its AdS/CFT dual
given by a non-linear $\sigma$-model \cite{Bena:2003wd,Beisert:2005bm}. Nowadays this is a wide area of research
where the integrability in the form of spin chain models and Bethe Ansatz plays the crucial role.
However, during the last year several new connections to the integrable world have emerged.

First, Nekrasov and Shatashvili discovered  \cite{Nekrasov:2009uh,Nekrasov:2009rc}
that $N=2$ supersymmetric gauge theories are in one to one correspondence with integrable Hamiltonian systems
in such a way that the supersymmetric vacua of gauge theories are mapped to Bethe states of the integrable models.
Thus, their identification heavily relies on Bethe Ansatz and its appearance in the description of BPS vacua.
Another interesting development has been done by Alday, Gaiotto and Tachikawa \cite{Alday:2009aq}
who related the instanton partition functions of certain $N=2$ SCFTs to conformal blocks of Liouville theory.
As was recently demonstrated in \cite{Nekrasov:2010ka}, the latter duality is actually closely related to
the former one.

On top of that, it was realized in \cite{Alday:2009dv} (see also \cite{Alday:2010vh,Hatsuda:2010cc})
that the classical problem of determining the minimal area surface in $AdS_5$ ending on a null polygonal contour,
which provides the strong coupling limit of gluon scattering amplitudes in $N=4$ super Yang-Mills \cite{Alday:2007hr},
has a solution in terms of Thermodynamic Bethe Ansatz (TBA) \cite{Zamolodchikov:1989cf}.
The TBA free energy then gives the area encoding the scattering amplitudes.

Here we consider another example of the unexpected interplay between gauge/string theory
and integrability whose physical explanation is still lacking.
Namely, it was noticed in \cite{Gaiotto:2008cd} (see also \cite{Gaiotto:2009hg})
that equations describing the exact moduli space of 4d $N=2$ supersymmetric gauge theory compactified
on a circle coincide with the equations of TBA.
Moreover, exactly the same equations appear also
in the twistor description of D-instanton corrected moduli space of Type II string theory
compactified on a Calabi-Yau (CY) threefold \cite{Alexandrov:2008gh,Alexandrov:2009zh}.
Due to this one may hope that the elaborated
machinery of integrable systems will give new insights for these research areas
as it did in the above mentioned situations.
However, for this one needs to extend the correspondence beyond just the level of equations
to some interesting physical quantities as well as to understand better
the integrable structure behind this TBA.

In this paper we try to fulfil this goal by pushing forward the relation between TBA and the instanton corrections
to the moduli spaces. In particular, we identify the TBA free energy with the instanton contribution
to the so called contact potential governing the quaternion-K\"ahler geometry of the string moduli space.
In the gauge theory context the analogous quantity is the K\"ahler potential. We show that its most non-trivial
part turns out to be encoded by the Yang--Yang functional \cite{Yang:1968rm} of the associated integrable system.

To uncover the integrable structure, we study the $S$-matrix which can be found from the TBA equations.
Although it is not unitary \cite{Gaiotto:2008cd}, we show that
it satisfies all standard constraints usually imposed on factorizable $S$-matrices of integrable systems.
They include Lorentz invariance, crossing symmetry, Yang-Baxter equation and bootstrap identity.
Finally, we derive the Y-system following from our TBA equations.

The paper is organized as follows. In the next section we review the twistor description of
hyperk\"ahler and quaternion-K\"ahler spaces which is crucial for introducing general instanton
corrections to the moduli spaces. These instanton corrections are further described
in section \ref{subsec_instcor} where we present the equations determining the non-perturbative geometry
of the moduli spaces and the results for the contact and K\"ahler potentials.
To our knowledge, the last quantity given in \eqref{Kahlerinst} was not known before.
The equations \eqref{eqXigi} represent our main starting point and the reader not interested
in their origin can pass directly to them.
In section \ref{sec_mainrel} we establish the relation of \eqref{eqXigi} to TBA and
express the contact and K\"ahler potentials in terms of the free energy and Yang--Yang functional, respectively.
After that in section \ref{sec_newint} we investigate the $S$-matrix associated to our problem,
derive the Y-system and comment on the conformal limit of our TBA.
In section \ref{sec_disc} we conclude with a discussion and some open questions.
In appendix \ref{ap_KP} we provide some details about evaluation and symplectic invariance of the K\"ahler potential.
Finally, appendix \ref{sec_example} is devoted to a particular case of rigid CY
where we observe the appearance of the MacMahon function \cite{Okounkov:2003sp}.

\section{Non-perturbative moduli spaces in gauge and string theories}

\subsection{Twistor description of HK and QK spaces}

The moduli spaces of compactified gauge and string theories with $N=2$ supersymmetry
are examples of manifolds with a quaternionic structure. The supersymmetry restricts
the moduli space to be hyperk\"ahler (HK) in gauge theory \cite{Seiberg:1996nz}
and its hypermultiplet sector to be quaternion-K\"ahler (QK) in string theory \cite{Bagger:1983tt}.
In both cases the geometry of these moduli spaces gets perturbative and non-perturbative contributions
and, of course, it is extremely difficult to describe them explicitly as corrections to the metric.
It would be much more convenient to have at our disposal some kind of prepotential, which
can encode all quantum corrections in a systematic way, similarly as does the holomorphic prepotential $F(X)$
for the vector multiplet moduli space. Thus, we are confronted with the problem of parametrizing
quaternionic geometries.

This problem is solved by considering the twistor space $\cZ_\cM$ of a quaternionic manifold $\cM$,
which is a $\CP$ bundle over $\cM$ and provides its very efficient description.
In particular, it allows to encode all geometric information in a set of holomorphic
functions playing the role of prepotentials we were asking for. Here we briefly
review the corresponding construction and refer to \cite{Alexandrov:2008ds,Alexandrov:2008nk} for more details.

The twistor description is so powerful because of the existence of a certain
holomorphic structure on $\cZ_\cM$. In the HK and QK cases, these structures are represented
locally by holomorphic forms, a 2-form $\Omega$ and a 1-form $\cX$, respectively.
The former defines a symplectic structure on each fiber of the bundle $\pi:\, \cZ_\cM\to\CP$, and the latter
produces a contact structure on $\cZ_\cM$.
As usual, locally it is always possible to choose Darboux coordinates where these forms take a standard form.
In other words, the twistor space can be covered by a set of open patches $\cU_i$ such that in each patch one has
\be
{\rm HK:}\quad \Omega^{[i]} =\de \mui{i}_\Lambda \wedge \de\nui{i}^\Lambda,
\qquad
{\rm QK:}\quad \cX^{[i]} =\de \ai{i} +\xii{i}^\Lambda \de\txii{i}_\Lambda,
\label{con1fo}
\ee
where $\Lambda=0,1,\dots,d-1$ with $4d=\dim_{\IR}\cM$, and
$(\nui{i}^\Lambda,\mui{i}_\Lambda,\zeta)$ (or $(\xii{i}^\Lambda,\txii{i}_\Lambda,\ai{i})$) form
a set of holomorphic Darboux coordinates in the patch $\cU_i$ of the twistor space of HK (respectively, QK) manifold.
Here $\zeta$ is a complex coordinate on $\CP$ (also in the QK case)
and we will use the notation $x^\mu$ to parameterize the base
quaternionic manifold $\cM$.

In addition to this holomorphic structure, the twistor space carries
a real structure defined in term of the antipodal map $\tau$
acting on $\CP$ as $\tau: \zeta\to-1/\bar\zeta$. To express the compatibility of the two
structures, we assume that the antipodal map sends $\cU_i$ to $\cU_{\bi}$.
Then the holomorphic forms should satisfy the reality constraint
\be
\overline{\tau(\Omega^{[i]})}=\Omega^{[\bi]},
\qquad
\overline{\tau(\hCX\ui{i})}= \hCX\ui{\bi},
\label{relcontact}
\ee
so that the Darboux coordinates may be chosen to satisfy similar relations, all with sign plus, under
the combined action of the complex conjugation and the antipodal map.

Something non-trivial appears when one considers the overlap of two patches $\cU_i\cap\cU_j$.
In this region the holomorphic forms defined in different patches must be related as
\be
\label{omij}
\Omega^{[i]}= f_{ij} ^2 \, \Omega^{[j]}  \  \mod\, \de\zeta\ui{i},
\qquad\qquad
\CX\ui{i} =  \hat f_{ij}^{2} \, \CX\ui{j},
\ee
where $f_{ij}$ are (fixed and simple) transition functions of the $\cO(1)$ bundle on $\CP$, whereas $\hat f_{ij}$
are fixed later. This implies that the local Darboux coordinates are related by symplectic
and contact transformations, respectively. Such transformations are generated by holomorphic functions on $\cZ_\cM$,
which we call transition functions $\Hij{ij}$. It is these functions that play the role of the prepotentials.
They establish relations between Darboux coordinates in different patches, which together with
the reality constraints and suitable boundary conditions allow to find these coordinates
as functions of $\zeta$ and $x^\mu$. These solutions, called twistor lines, contain all geometric information
and are the starting point to find the metrics on $\cM$ and its twistor space.

In the following we restrict our attention only to those transition functions
which in the QK case are independent on the coordinates $\ai{i}$.
Physically, for the hypermultiplet moduli space of compactified Type II string theory
this means that we ignore the contributions due to NS5-brane instantons \cite{Becker:1995kb}.
Such restriction brings great simplifications.
In particular, the coefficients $\hat f_{ij}$ left undetermined above are actually
given by
\be
\label{glue2}
\hat f_{ij}^2=1-\p_{\ai{j} }\hHij{ij}
\ee
and thus reduce to 1 in this case. What is important for us is that given this restriction
the equations determining the twistor lines in HK and QK cases become equivalent.
Indeed, in the QK case for $\alpha$-independent $\Hij{ij}$ they look as
\be
\xii{j}^\Lambda =  \xii{i}^\Lambda -\p_{\txii{j}_\Lambda }\hHij{ij},
\qquad
\txii{j}_\Lambda =  \txii{i}_\Lambda
 + \p_{\xii{i}^\Lambda } \hHij{ij},
\label{QKgluing}
\ee
whereas in the HK case they are the same provided one replaces $\xii{i}^\Lambda$ by
$\etai{i}^\Lambda\equiv (\I\zeta)^{-1} f_{+i}^2\nui{i}^\Lambda$
and $\txii{i}_\Lambda$ by $\mui{i}_\Lambda$. Here the index $\scriptstyle +$
refers to some fixed patch $\cU_+$ which we choose to be the one around
the north pole $\zeta=0$. The coincidence of the equations allows us to consider the HK and QK cases
simultaneously. The notations mostly used in the following correspond to the QK case.

The gluing conditions \eqref{QKgluing} for the twistor lines
can be rewritten in a form more convenient for a perturbative treatment as the following
integral equations \cite{Alexandrov:2009zh}\footnote{Comparing to \cite{Gaiotto:2008cd,Alexandrov:2008gh},
we changed some normalizations to avoid some numerical factors and to make the symplectic invariance more explicit.
In particular, $\cX,\mu_\Lambda, \txi_\Lambda, \alpha$
and $\Hij{ij}$ are renormalized by factor $-2\I$, $\Omega,e^\phi$ by 2, and $\nu^\Lambda$ by $\I$.
}
\bea
\xii{i}^\Lambda(\varpi,x^\mu)& =& A^\Lambda +
\varpi^{-1} Y^\Lambda - \varpi \bY^\Lambda
-\frac12 \sum_j \oint_{C_j}\frac{\de\varpi'}{2\pi\I \varpi'}\,
\frac{\varpi'+\varpi}{\varpi'-\varpi}\, \p_{\txii{j}_\Lambda }\hHij{ij}(\varpi') ,
\nonumber \\
\txi_\Lambda^{[i]}(\varpi,x^\mu)& = &B_\Lambda +
\half  \sum_j \oint_{C_j} \frac{\de \varpi'}{2 \pi \I \varpi'} \,
\frac{\varpi' + \varpi}{\varpi' - \varpi}
\, \p_{\xii{i}^\Lambda } \hHij{ij}(\varpi'),
\label{txiqline}
\eea
where $\zeta\in \cU_i$, $C_j$ is the contour surrounding $\cU_j$ in the counterclockwise direction, whereas
complex $Y^\Lambda$ and real $A^\Lambda, B_\Lambda$ are free parameters playing the role
of coordinates on $\cM$.\footnote{Actually in the QK case the phase of $Y^0$ can be absorbed into $\zeta$
so that it becomes real. The missing coordinate is provided by the real part of the constant coefficient
$B_\alpha$ in the expansion of $\ai{+}$, which we do not write here explicitly.}
The sum in \eqref{txiqline} goes over all patches including those
which do not intersect with $\cU_i$. In that case the transition functions are defined by the
cocycle condition and by analytic continuation \cite{Alexandrov:2008ds}.

Once these integral equations are solved, there is a straightforward procedure
to extract the metric \cite{Alexandrov:2008nk,Alexandrov:2008gh}.
On this way the prominent role is played by the K\"ahler potential $K_\cM$
in the HK case and the so called contact potential $\Phi\di{i}$ in the QK case,
which provides a \kahler potential for the metric on the twistor space $\cZ_\cM$
\be
\label{Knuflat}
K_{\cZ}\ui{i} = \log\frac{1+\varpi\bar \varpi}{|\varpi|}
+ \Re\Phi\di{i} .
\ee
The contact potential appears as a certain coefficient in the expansion of the contact one-form
\be
\label{contact}
\hCX\ui{i} = 2\,  \frac{e^{\Phi\di{i}}}{\I\varpi}
\(\de\varpi + p_+ -\I p_3 \,\varpi + p_-\, \varpi^2\) ,
\ee
where $\vec p$ is the $SU(2)$ part of the Levi-Civita connection on $\cM$,
and generically it is a function $\Phi\di{i}(x^\mu,\zeta)$ holomorphic on the $\CP$ fiber, but
defined only locally what is reflected by the patch index.
However, in the case under consideration when the transition functions are $\alpha$-independent,
it becomes real, globally defined and independent on $\zeta$, $\Phi\di{i}=\phi(x^\mu)$.
The remaining function on the base manifold is given by
\be
e^\phi=\frac{1}{8\pi} \sum_j\oint_{C_j}\frac{\de\varpi}{\varpi}
\(\varpi^{-1} Y^{\Lambda}-\varpi \bY^{\Lambda} \)
\p_{\xii{i}^\Lambda } \hHij{ij}
+\cij{+}_\alpha,
\label{contpotconst}
\ee
where $c_\alpha$ is a constant called anomalous dimension, which encodes a boundary condition
for the twistor line $\ai{+}$.

In the HK case the K\"ahler potential can also be expressed as an integral of transition functions.
In our notations the representation found in \cite{Alexandrov:2009zh} can be summarized as
\be
\label{Kdef}
K_\cM= \frac{1}{4\pi}\sum_j\oint_{C_j} \frac{\de\zeta}{\zeta}
\[\Hij{ij}-\(\mui{j}_\Lambda-B_\Lambda\)\p_{\mui{j}_\Lambda}\Hij{ij}
-A^\Lambda\p_{\etai{i}^\Lambda} \Hij{ij} \].
\ee
By simple manipulations using the integral equations \eqref{txiqline}, it can be rewritten
in the following form
\be
\label{Kdefrew}
K_\cM= \frac{1}{4\pi}\sum_j\oint_{C_j} \frac{\de\zeta}{\zeta}
\[\Hij{ij}-\etai{i}^\Lambda \p_{\etai{i}^\Lambda}\Hij{ij}
+\(\varpi^{-1} Y^{\Lambda}-\varpi \bY^{\Lambda} \)\p_{\etai{i}^\Lambda} \Hij{ij} \].
\ee
This shows that in the physically important case of transition functions homogeneous of degree one,
relevant when the HK space is the hyperk\"ahler cone of a QK space \cite{deWit:2001dj,Alexandrov:2008nk},
the K\"ahler potential essentially coincides with the contact potential \eqref{contpotconst}.

We close this subsection by noting that although all integration contours $C_j$
appearing in the formulae above are closed since they surround open patches,
under some conditions it is possible to generalize these results to include also open contours.
Typically, such open contours can be viewed as a leftover of some discontinuity cuts
in transition functions of a ``more fundamental" construction using only open patches
and closed contours. The former arises from the latter when one
shrinks the contour surrounding the cut so that its contribution reduces to an integral of the discontinuity
of the initial integrand.
Such description in terms of open contours turns out to be relevant in the
discussion of the instanton corrected moduli spaces in the next subsection.

\subsection{Instanton contributions to the moduli spaces}
\label{subsec_instcor}

The low energy dynamics of $N=2$ supersymmetric gauge and string theories is completely determined
by the geometry of their moduli spaces. For a 4d $N=2$ gauge theory with the gauge group $G$ of rank $d$,
which is compactified on a circle of radius $R$,
the moduli space is parameterized by complex scalars $z^\Lambda$ from the vector multiplet
and by Wilson lines of the gauge potential around the circle, which have ``electric" $\zeta^\Lambda$
and ``magnetic" components $\tzeta_{\Lambda}$ and are all periodic.\footnote{We use different notations
than those of \cite{Gaiotto:2008cd}. The fields are denoted by the same letters which are used
to parametrize the moduli space of Type IIA string theory since in all equations they appear in exactly the same way.
Besides, the Wilson lines are normalized to have period 1.}
It is a hyperk\"ahler manifold whose perturbative metric follows from the simple 3d truncation of the
4d vector multiplet Lagrangian and thus is completely defined by the holomorphic prepotential $F(z)$.
However, it gets instanton contributions from the massive spectrum due to
BPS particles going around the compactification circle. For large $R$ these contributions are exponentially suppressed
since for a particle of charge $\gamma=(q_\Lambda,p^\Lambda)$ they are weighted
by $e^{-2\pi R |Z_\gamma|}$ where
\be
\label{defZ}
Z_\gamma(z) = q_\Lambda z^\Lambda- p^\Lambda F_\Lambda(z)
\ee
is the central charge function giving the mass of the BPS particle.

The low energy physics of Type IIA string theory compactified on
a CY threefold $X$ is a bit different. Its complete moduli space is factorized
to the moduli spaces of vector and hypermultiplets.
The former is tree level exact so that our interest is concentrated on the hypermultiplet sector.
It comprises $d=h_{2,1}(X)+1$ hypermultiplets, which include
the complex structure moduli $X^\Lambda=\int_{\gamma^\Lambda} \Omega$,
$F_\Lambda=\int_{\gamma_\Lambda} \Omega$, the RR scalars $\zeta^\Lambda, \tzeta_\Lambda$
representing the RR three-form integrated along a symplectic basis $(\gamma^\Lambda,\gamma_\Lambda)$
of A and B cycles in $H_3(X,\IZ)$, the four-dimensional dilaton
$e^{\phi}=1/g_{(4)}^2$ and the Neveu-Schwarz (NS) axion $\sigma$, dual to the
NS two-form $B$ in four dimensions. Whereas $X^\Lambda$ provide a set
of homogeneous coordinates for complex structure deformations,
they may be traded for the inhomogeneous coordinates $z^a=X^a/X^0$.
As we mentioned above, this moduli space is a \qk manifold.
Its tree level metric is determined by the holomorphic prepotential $F(z)$ as in gauge theory
and receives one-loop \cite{Antoniadis:1997eg,Robles-Llana:2006ez}
and instanton corrections \cite{Becker:1995kb}. The latter arise either from D2-branes
wrapping non-trivial 3-dimensional cycles of $X$ or from the NS5-brane wrapping the whole Calabi-Yau.
In this work we ignore the second type of instantons so that we remain only with membrane contributions.
Since cycles in $H_3(X,\IZ)$ are parameterized by the symplectic vector $\gamma=(q_\Lambda,p^\Lambda)$
and the weight of the corresponding instanton is determined by the same function \eqref{defZ} (with $z^0\equiv 1$),
one may expect that the D-instanton corrected hypermultiplet moduli space and the exact gauge theory moduli space
from above have a similar description.

This indeed turns out to be the case. Such description is provided by the twistor formalism
from the previous subsection, which as we have seen works in the same way for HK and QK spaces.
Below we review the corresponding construction for the gauge and string moduli spaces \cite{Gaiotto:2008cd,Alexandrov:2008gh}.
It essentially amounts to provide a covering of $\CP$ and a set of transition functions.
First, we present it for the moduli spaces at the perturbative level and then show how the instantons are included.

To construct the perturbative moduli space,
let us cover the Riemann sphere by the following three patches:
the first patch $\cU_+$ surrounds the north pole, the second patch $\cU_-$ surrounds
the south pole and the rest is covered by $\cU_0$ (see Fig. \ref{figcover}a).
There are two non-trivial transition functions associated with such covering.
\be
\label{gensymp}
\hHij{+0}=   F(\xii{+}),
\qquad
\hHij{-0} = \bF(\xii{-}).
\ee
As usual, in the gauge theory case one should replace $\xii{i}$ by $\etai{i}$.\footnote{In fact, in the gauge
theory case the holomorphic prepotential contains a logarithmic term and thus it is only a {\it quasi}-homogeneous
function of second degree
\be
z^\Lambda F_\Lambda=2F+\hf \, Q_{\Lambda\Sigma}z^\Lambda z^\Sigma,
\label{quasiF}
\ee
where $Q_{\Lambda\Sigma}$ is a matrix constructed from charges of hypermultiplets
whose explicit form will not be important in the following. On the other hand, the tree level
transition functions should not contain logarithmic singularities. Due to this, the functions \eqref{gensymp}
should be actually replaced by
\be
\label{gensympG}
\hHij{+0}=   \zeta^{-2} F(\zeta\etai{+}),
\qquad
\hHij{-0} = \zeta^2\bF(\zeta^{-1}\etai{-}).
\ee
}
On the other hand, in the string theory case one should add information about the anomalous dimension
$c_\alpha$ (see \eqref{contpotconst}). It turns out to be determined by the Euler number of CY,
$\ci{+}_\alpha= \chi_X/(96\pi)$, and incorporates the one-loop correction \cite{Alexandrov:2008nk}.
Then the equations \eqref{txiqline} provide an explicit representation
for the twistor lines. Identifying properly the abstract coordinates $Z^\Lambda,A^\Lambda,B_\Lambda$ of the twistor approach
with the physical fields, in the patch $\cU_0$ (in string notations) one finds
\be
\begin{split}
\xii{0}^\Lambda &=\xip^\Lambda\equiv \zeta^\Lambda + \cR
\left( \varpi^{-1} z^{\Lambda} -\varpi \,\bz^{\Lambda}  \right),
\\
\txii{0}_\Lambda &= \txip_\Lambda\equiv \tzeta_\Lambda + \cR
\left( \varpi^{-1} F_\Lambda(z)-\varpi \,\bF_\Lambda(\bz) \right),
\end{split}
\label{sftwistl}
\ee
where $\cR=R/2$ in the gauge theory case and can be expressed through the dilaton, using
\be
\label{phipertA}
e^{\phi}= \frac{\cR^2}{2}\,K(z,\bz)+ \frac{\chi_X}{96\pi}
\ee
with $K(z,\bz)\equiv-2 \Im(\bz^\Lambda F_\Lambda)$, in string theory.
Remarkably, the contact potential \eqref{contpotconst} turns out to coincide with the dilaton, which is in turn
related to the four-dimensional string coupling $g_{(4)}$. On the other hand, $\cR$ is inversely proportional to the ten-dimensional
string coupling $g_{(10)}$. These identifications will survive the instanton corrections.
Altogether the above results are sufficient to extract the metric
in a straightforward way \cite{Alexandrov:2008nk}.

\lfig{Covering of $\CP$ and transition functions of the perturbative
and instanton corrected twistor spaces.}{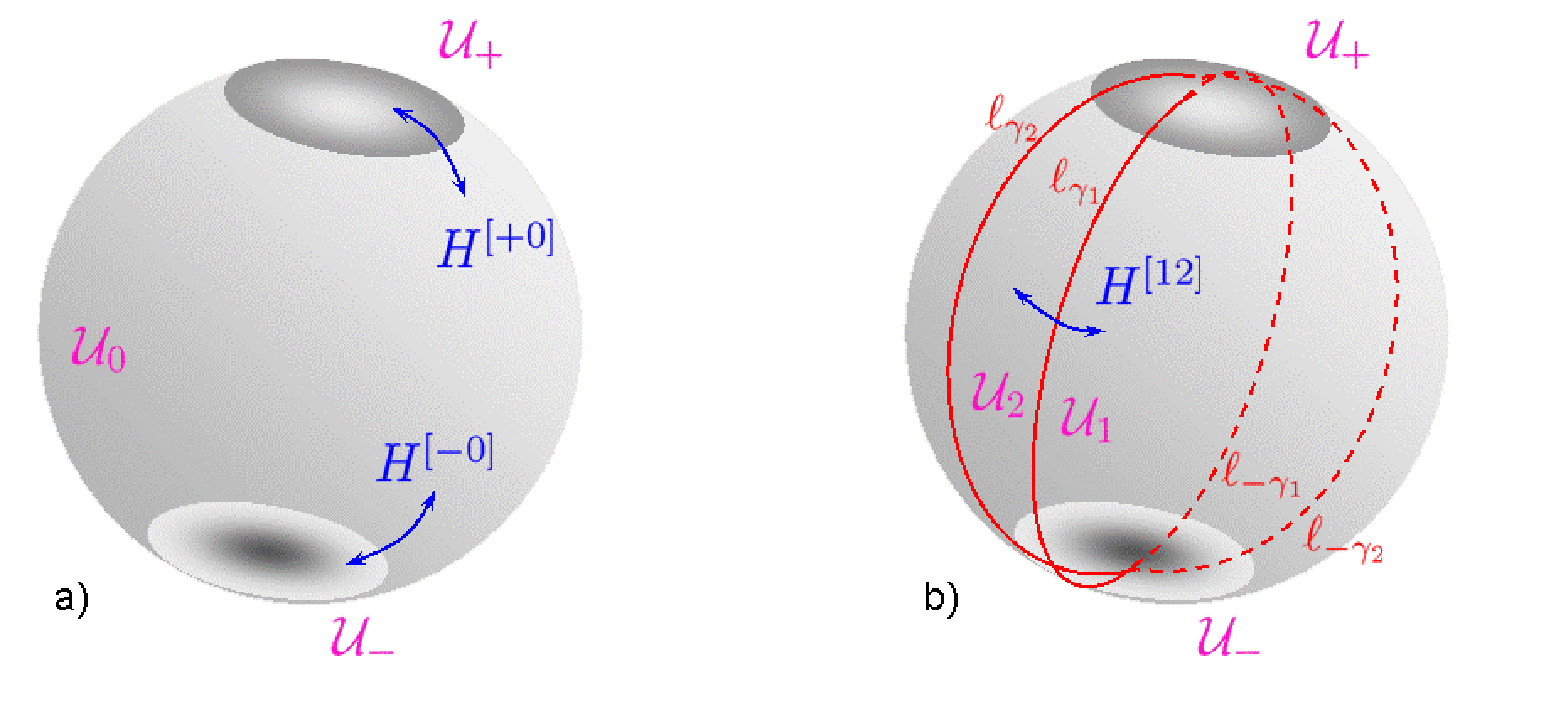}{7.5cm}{figcover}

Let us include the instanton contributions from a set of $2N$ BPS particles
(or D2-branes) with charges $\{ \gamma_a\}_{a=1}^{2N}$. The number of particles is taken even
because every particle is accompanied by its antiparticle with the opposite charge.
Eventually we are interested in the limit $N\to\infty$. We assume that the charges are ordered
in accordance with decreasing of the phase of $Z_{\gamma}$ \eqref{defZ} and
that these phases for the charges $\gamma_a$ which are not mutually local, {\it i.e.}
for those which have non-vanishing symplectic invariant scalar product
\be
\<\gamma_1,\gamma_2\>=q_{1,\Lambda}p_2^\Lambda-q_{2,\Lambda}p_1^\Lambda,
\label{scpr}
\ee
are all different. For mutually local charges with coinciding phases of $Z_{\gamma}$
the order is not important.

Each charge vector $\gamma$ defines a ``BPS ray''
$\ellg{\gamma}$ on $\CP$ going between the north and south poles as
\be
\ellg{\gamma}= \{ \varpi :\,  Z(\gamma)/\varpi \in \I\IR^{-} \} .
\label{rays}
\ee
These rays split the patch $\cU_0$ into $2N$ sectors which we call $\cU_a$,
so that the covering of $\CP$ consists now of these connected parts
divided by the contours $\ellg{\gamma_a}$ and the usual patches $\cU_\pm$ (see Fig. \ref{figcover}b).

What we have to provide is the transition function through the BPS ray associated with a charge $\gamma_a$.
For that purpose we define
\be
\Xiijg{ab}{c} \equiv
q_{c,\Lambda}\xii{a}^\Lambda- p_c^\Lambda \txii{b}_\Lambda.
\label{Xarggem}
\ee
Then the transition function through the BPS ray is \cite{Alexandrov:2009zh}
\be
\hHij{a\,a+1}(\xii{a},\txii{a+1})
=\Ggss{\gamma_a}-\hf\,q_{a,\Lambda} p_a^\Lambda (\Ggss{\gamma_a}')^2 ,
\label{transellg}
\ee
where $\Ggss{\gamma_a}(\Xigi{a})$ with $\Xigi{a}\equiv\Xiijg{aa}{a}$ is defined in term of
the dilogarithm function as
\be
\Ggs(\Xikl)=\frac{\hnkl}{(2\pi)^2}\,
\Li_2\left(e^{-2\pi \I \Xikl} \right).
\label{prepH}
\ee
The coefficient $\hnkl$ encodes the spectrum of the gauge theory or carries a topological information
about Calabi-Yau. In the former case it was identified with the second helicity supertrace in \cite{Gaiotto:2008cd} and
in general it can be related to the generalized Donaldson--Thomas invariants introduced in \cite{ks}.
For us two facts are important: $\hnkl$ are integer numbers and do not depend on the sign of the charge,
$n_\gamma=n_{-\gamma}$.
The presence of the second (non-symplectic invariant) term in \eqref{transellg} is related to the fact
that the arguments of transition functions should be Darboux coordinates from different patches.
Therefore, the r.h.s. of \eqref{transellg} should be expressed through $\txii{a+1}_\Lambda$. This is achieved by solving
\be
\Xiijg{a\,a+1}{a}
=\Xigi{a} - q_{a,\Lambda} p_a^\Lambda \Ggss{\gamma_a}'(\Xigi{a})
\label{Xargsim}
\ee
for $\Xigi{a}$ which, of course, cannot be done explicitly.
Nevertheless, this is sufficient to define the derivatives of $\Hij{a\,a+1}$
which lead to the following simple symplectic transformations
\be
\xii{a+1}^\Lambda=\xii{a}^\Lambda+p_a^\Lambda\Ggss{\gamma_a}',
\qquad
\txii{a+1}_\Lambda=\txii{a}_\Lambda+q_{a,\Lambda}\Ggss{\gamma_a}'.
\label{glumany}
\ee
To complete the construction we need also to provide $\Hij{\pm a}$.
They are given by \eqref{gensymp} plus instanton contributions which
are known but quite complicated. We refer to \cite{Alexandrov:2009zh}
for their explicit expressions.

Note that this construction realizes the option mentioned in the end of the previous subsection.
Namely, the transition functions \eqref{transellg} are to be integrated along open contours given by the BPS rays.
There exists also a version with closed contours \cite{Alexandrov:2008gh}, which however does not bring anything new.

Another useful comment is that this picture gives an interesting geometric interpretation
of the wall crossing phenomenon. The lines of marginal stability where it takes place arise at the moduli configurations
where the phases of the central charge \eqref{defZ} of two BPS states align. This happens precisely
when the two BPS rays \eqref{rays} cross each other. Since in general different order of BPS rays leads
to inequivalent constructions, this is a clear origin of discontinuities. On the other hand,
crossing a line of marginal stability, the BPS spectrum, encoded in the coefficients $\hnkl$, also changes discontinuously.
A formula recently found by Kontsevich and Soibelman \cite{ks} ensures that the two effects compensate each other
and the resulting metric is continuous \cite{Gaiotto:2008cd}.

Although the presented construction uniquely defines the twistor space and the underlying quaternionic
manifold, to actually compute the metric one needs to solve the integral equations \eqref{txiqline}
for the twistor lines.
In our case, they can be reduced to the following system\footnote{Here we ignore an additional
issue of sign in front of the exponent in the logarithm appearing in the gauge theory context.
It is related to some subtleties in the fermion number of bound states and can be taken into account by
the so called ``quadratic refinement" \cite{Gaiotto:2008cd}. It affects the solution in a simple way
and is not important for our discussion.\label{foot_quadrref}}
{\renewcommand{\arraystretch}{0}
\be
\label{eqXigi}
\begin{array}{|c|}
\hline\rule{0pt}{3pt}
\\
 \rule{0pt}{3pt}
\displaystyle\
\Xigi{a}(\varpi)=\Thkli{a}+\cR\(\varpi^{-1}Z_{\gamma_a}-\varpi\bZ_{\gamma_a}\)+
\frac{1}{8\pi^2}\sum_{b\ne a} \hng{b}\<\gamma_a,\gamma_b\> \int_{\ellg{\gamma_b}}\frac{\d \varpi'}{\varpi'}\,
\frac{\varpi+\varpi'}{\varpi-\varpi'}\,
\log\(1-e^{-2\pi \I \Xigi{b}(\varpi')}\)
\\
 \rule{0pt}{3pt} \\
\hline
\end{array}
\ee
}
where
\be
\Thkl
\equiv q_\Lambda \zeta^\Lambda - p^\Lambda\tzeta_\Lambda.
\label{THg}
\ee
These equations encode all non-trivialities of the problem.
For large $\cR$ (large circle radius or small 10d string coupling),
they can be analyzed perturbatively and their solution, represented by the set of variables $\Xigi{a}(\varpi)$,
contains all orders of the instanton expansion.
Note that due to the reality conditions on twistor lines and the fact that $\cU_{\ba}$ is the patch
associated with the charge $\gamma_{\ba}=-\gamma_a$, the functions $\Xigi{a}(\varpi)$
satisfy
{\renewcommand{\arraystretch}{0}
\be
\label{realXig}
\begin{array}{|c|}
\hline\rule{0pt}{3pt}
\\
 \rule{0pt}{3pt}
\displaystyle\
\overline{\Xigi{a}^{\vphantom{A}}}(\varpi)=-\Xigmi{a}\(-\varpi^{-1}\).
\vphantom{\frac{A}{A}}
\\
 \rule{0pt}{3pt} \\
\hline
\end{array}
\ee
}

Once the functions $\Xigi{a}(\varpi)$ are known, the twistor lines easily follow as \cite{Alexandrov:2009zh}
\be
\label{exline}
\begin{split}
\xii{a}^\Lambda &= \xip^\Lambda +
\frac{1}{8\pi^2}\sum_b \hng{b} p_b^\Lambda \Igg{b}(\varpi) ,
\\
\txii{a}_\Lambda &=
\txip_\Lambda
+\frac{1}{8\pi^2}\sum_b \hng{b} q_{b,\Lambda}\Igg{b}(\varpi)  ,
\end{split}
\ee
where $\varpi\in \cU_a$ and
\be
\Igg{}(\varpi)=\int_{\ellg{\gamma}}\frac{\d \varpi'}{\varpi'}\,
\frac{\varpi+\varpi'}{\varpi-\varpi'}\,
\log\(1-e^{-2\pi \I \Xigi{}(\varpi')}\).
\label{newfun}
\ee
The general formula for the contact potential \eqref{contpotconst}
gives the following result
\be
e^{\phi} = \frac{\cR^2}{2}\, K(z,\bz)+\frac{\chi_X}{96\pi}
-\frac{ \I\cR}{16\pi^2}\sum\limits_{a} \hng{a}
\int_{\ellg{\gamma_a}}\frac{\d \varpi}{\varpi}\,
\( \varpi^{-1}Z_{\gamma_a} -\varpi\bZ_{\gamma_a}\)
\log\(1-e^{-2\pi \I \Xigi{a}(\varpi)}\)
\label{phiinstmany}
\ee
and can be used to trade $\cR$ for the dilaton.
Similarly, one can get an explicit expression for the K\"ahler potential \eqref{Kdefrew},
which to our knowledge did not appear so far in the literature. Some details of its
derivation can be found in Appendix \ref{ap_KP}. Here we just present the final
result\footnote{In our normalization the K\"ahler potential differs from the one in \cite{Seiberg:1996nz,Gaiotto:2008cd}
by a constant $R$-dependent factor.}
\bea
K_\cM&= & \frac{\cR^2}{2}\, K(z,\bz)-\hf\, \Im F_{\Lambda\Sigma}\(\zeta^\Lambda\zeta^\Sigma+
\frac{1}{64\pi^4}\,\sum_{a,b}\hng{a}\hng{b}p_a^\Lambda p_b^\Sigma
\int_{\ellg{\gamma_a}}\!\!\cD_a\zeta
\int_{\ellg{\gamma_b}}\!\!\cD_b\zeta' \)
\nn\\
&& -\frac{1}{16\pi^3}\, \sum_{a}\hng{a}\int_{\ellg{\gamma_a}}\!\! \frac{\d \varpi}{\varpi}
\[\Li_2 \(e^{-2\pi \I \Xigi{a}}\)
-2\pi \I \zeta^\Lambda\(q_{a,\Lambda}-p_a^\Sigma\Re F_{\Lambda\Sigma}\)\log\(1-e^{-2\pi \I \Xigi{a}}\)\]
\nn\\
&& +\frac{\I}{128\pi^4 }\, \sum_{a\ne b} \hng{a}\hng{b}\<\gamma_a,\gamma_b\>
\int_{\ellg{\gamma_a}}\!\!\cD_a\zeta \int_{\ellg{\gamma_b}}\!\!\cD_b\zeta'\,
\frac{\varpi+\varpi'}{\varpi-\varpi'},
\label{Kahlerinst}
\eea
where we abbreviated
\be
\cD_a\zeta=\frac{\d \varpi}{\varpi}\,\log\(1-e^{-2\pi \I \Xigi{a}(\varpi)}\).
\label{measure}
\ee
As we shall see, both quantities, \eqref{phiinstmany} and \eqref{Kahlerinst}, appear naturally
also in the context of TBA in the next section.
Besides, note that since $\Xigi{a}$ is symplectic invariant, the twistor lines \eqref{exline}
form a vector under symplectic transformations, the contact potential is invariant,
whereas the K\"ahler potential can be shown to be invariant up to a K\"ahler transformation (see Appendix \ref{ap_KP}).
Thus, the whole construction respects the symplectic symmetry.

\section{Moduli space geometry and TBA}
\label{sec_mainrel}

\subsection{Relation to TBA}
\label{subsec_releq}

As was noticed in \cite{Gaiotto:2008cd} (Appendix E of the revised version),
the equations \eqref{eqXigi}, which encode the geometry of
the instanton corrected moduli space, turn out to coincide with the equations of TBA.
These are equations for an integrable system of particles in $1+1$ dimensions. The particles
are characterized by spectral densities $\epsilon_a(\theta)$ considered
as functions of the rapidity parameter $\theta$ which defines their two-dimensional momentum.
TBA imposes the following non-linear integral equations on the spectral densities \cite{Zamolodchikov:1989cf}:
\be
m_a \beta\cosh\theta=\epsilon_a(\theta)+\frac{1}{2\pi}\sum_{b}\int_{-\infty}^\infty\d \theta'\,
\phi_{ab}(\theta-\theta')\log\(1+e^{\beta\mu_b-\epsilon_b(\theta')}\),
\label{Bethe}
\ee
where $\beta$ is the inverse temperature, $m_a$ are mass parameters for the particles of type $a$,
$\mu_a$ are their chemical potentials,
and $\phi_{ab}(\theta)=-\I\frac{\p\log S_{ab}}{\p\theta}$ is defined by the two-particle $S$-matrix $S_{ab}(\theta)$.

To establish a relation between \eqref{eqXigi} and \eqref{Bethe}, we set
\be
\epsilon_a(\theta)=2\pi \I \(\Xigi{a}(\I e^{\I \psi_a +\theta})-\Theta_a\),
\ee
where $\psi_a=\arg Z_{\gamma_a}$. This implies that for every charge one changes the coordinate on $\CP$ as
$\zeta=\I e^{\I \psi_a +\theta}$ so that the BPS ray $\ellg{\gamma_a}$ is mapped to the real axis.
Then plugging these changes into \eqref{eqXigi}, one obtains
\be
\label{eqYigi}
4\pi \cR|Z_{\gamma_a}|\cosh\theta=\epsilon_a(\theta)-
\frac{\I}{4\pi}\sum_{b\ne a} \hng{b}\<\gamma_a,\gamma_b\> \int\limits_{-\infty}^\infty {\d \theta'}\,
\frac{e^{\theta-\theta'}+\Psi_{ab}}{e^{\theta-\theta'}-\Psi_{ab}}\,
\log\(1-e^{-2\pi \I \Theta_b-\epsilon_b(\theta')}\),
\ee
where $\Psi_{ab}=e^{\I \(\psi_{b}-\psi_a\)}$.
Comparing \eqref{eqYigi} with \eqref{Bethe}, one finds that they have the same form if one identifies
\be
\beta m_a =4\pi \cR|Z_{\gamma_a}|,
\qquad
\beta\mu_a=-2\pi \I \Theta_b+\pi \I,
\qquad
\phi_{ab}(\theta)=-\frac{\I}{2}\,\<\gamma_a,\gamma_b\>\frac{e^{\theta}+\Psi_{ab}}{e^{\theta}-\Psi_{ab}},
\label{ident}
\ee
whereas the additional factor $\hng{b}$ in the sum is considered as a weight of the particles of type $b$.

\subsection{Potentials, free energy and Yang--Yang functional}
\label{subsec_relpot}

The curious relation observed in \cite{Gaiotto:2008cd} and presented in the previous subsection
can be deepen by considering some quantities playing an important role in TBA and comparing them with
potentials of quaternionic geometries of the moduli spaces, the contact potential
\eqref{phiinstmany} and the K\"ahler potential \eqref{Kahlerinst}.

First, the most important quantity, which is usually considered in the context of TBA, is the free energy
of the integrable system. It is given by
\be
\cF(\beta)=\frac{\beta}{2\pi}\sum_{a} m_a \int_{-\infty}^\infty\d \theta\,
\cosh\theta\,\log\(1+e^{\beta\mu_a-\epsilon_a(\theta)}\).
\label{grstener}
\ee
It is straightforward to check that it coincides with
the instanton part of the contact potential
so that one has the following relation
{\renewcommand{\arraystretch}{0}
\be
\label{identener}
\begin{array}{|c|}
\hline\rule{0pt}{3pt}
\\
 \rule{0pt}{3pt}
\displaystyle\
e^{\phi}=  \frac{\cR^2}{2}\, K(z,\bz)+\frac{\chi_X}{96\pi}-\frac{1}{16\pi^2}\,\cF(\beta).
\vphantom{\frac{A^A}{A_A}}
\\
 \rule{0pt}{3pt} \\
\hline
\end{array}
\ee
}

On the other hand, the K\"ahler potential of the gauge theory turns out
to be related to the so called Yang--Yang functional \cite{Yang:1968rm}. This is a functional
which generates the action principle for the Bethe equations following
from it by varying with respect to the spectral densities.
It can be conveniently written as \cite{Nekrasov:2009rc}
\be
\cW[\vph,\rho]=\frac{1}{8\pi^2}\sum_{a,b}\int \de\theta\int \de\theta' \,
\phi_{ab}(\theta-\theta')\rho_a(\theta)\rho_b(\theta')
+\frac{1}{2\pi}\sum_a \int\de \theta \[\rho_a(\theta)\vph_a(\theta)-\Li_2\(e^{\lambda_a(\theta)-\vph_a(\theta)}\)\].
\label{YYfun}
\ee
Here $\vph_a(\theta)$ is the interacting part of the spectral density and $\lambda_a(\theta)$ encodes
its free part together with the chemical potential
\be
\vph_a(\theta)=\eps_a(\theta)-m_a\beta\cosh\theta,
\qquad
\lambda_a(\theta) =\beta(\mu_a-m_a\cosh\theta)-\pi\I.
\label{identYY}
\ee
Varying \eqref{YYfun} with respect to $\vph_a$ and $\rho_a$ and using \eqref{identYY},
one indeed gets the equations of TBA \eqref{Bethe}.
The critical value of the Yang--Yang functional is then given by
\be
\begin{split}
\cW_{\rm cr}=&
-\frac{1}{8\pi^2}\sum_{a,b}\int \de\theta\int \de\theta' \,
\phi_{ab}(\theta-\theta')\log\(1-e^{\lambda_a(\theta)-\vph_a(\theta)}\)\log\(1-e^{\lambda_b(\theta')-\vph_b(\theta')}\)
\\
& -\frac{1}{2\pi}\sum_a \int\de \theta\, \Li_2\(e^{\lambda_a(\theta)-\vph_a(\theta)}\).
\end{split}
\label{YYfuncrit}
\ee
Comparison of this expression with the exact K\"ahler potential \eqref{Kahlerinst} of the gauge
theory moduli space reveals that it reproduces two instanton symplectic invariant terms of the latter.
Thus, one has the following relation
{\renewcommand{\arraystretch}{0}
\be
\label{relYYK}
\begin{array}{|c|}
\hline\rule{0pt}{3pt}
\\
 \rule{0pt}{3pt}
\displaystyle\
\begin{split}
\vphantom{\frac{A^A}{A_A}}
K_\cM= & \frac{\cR^2}{2}\, K(z,\bz)
-\frac{1}{4}\, N^{\Lambda\Sigma}(w_\Lambda-\bw_\Lambda)(w_\Sigma-\bw_\Sigma)
\\
&
+\frac{1}{64\pi^4}\,\sum_{a,b}\hng{a}\hng{b}\cQ_{ab}
\int_{\ellg{\gamma_a}}\!\!\cD_a\zeta
\int_{\ellg{\gamma_b}}\!\!\cD_b\zeta'
+\frac{1}{8\pi^2}\,\cW_{\rm cr},
\vphantom{\frac{A^A}{A_A}}
\end{split}
\\
 \rule{0pt}{3pt} \\
\hline
\end{array}
\ee
}
where $N^{\Lambda\Sigma}$ is the inverse of $N_{\Lambda\Sigma}=-2\Im F_{\Lambda\Sigma}$, $\cQ_{ab}$ is constructed
in terms of charges as
\be
\cQ_{ab}=\frac{1}{4}\, N_{\Lambda\Sigma}p_a^\Lambda p_b^\Sigma
+N^{\Lambda\Sigma}\(q_{a,\Lambda}-p_a^\Theta\Re F_{\Lambda\Theta}\)\(q_{b,\Sigma}-p_b^\Xi\Re F_{\Sigma\Xi}\),
\ee
and we used the holomorphic coordinates $w_\Lambda$ defined in \eqref{defholw}.
Although we found that the Yang--Yang functional does not coincide with the full instanton contribution to $K_\cM$,
it captures the most non-trivial part of the K\"ahler potential. In general, since $K_\cM$
is subject to K\"ahler transformations, the exact equality should not be expected and
the appearance of the additional terms is not surprising.

\section{Integrable structure of instanton contributions}
\label{sec_newint}

\subsection{The $S$-matrix}
\label{subsec_Smat}

The $S$-matrix corresponding to TBA \eqref{eqYigi} can be easily obtained by integrating
$\phi_{ab}(\theta)$ from \eqref{ident}. In this way, one finds
\be
S_{ab}(\theta)=C_{ab}\[\sinh\(\hf\(\theta+\I(\psi_a-\psi_b)\)\)
\]^{\<\gamma_a,\gamma_b\>},
\label{Smat}
\ee
where $C_{ab}$ is an integration constant. As was noticed in \cite{Gaiotto:2008cd},
this $S$-matrix is non-unitary. However, it is not necessarily a problem
since nowadays there are many non-unitary integrable models.
On the other hand, integrability and consistent physical interpretation require
the $S$-matrix to satisfy a set of severe conditions.
In this section we show that our $S$-matrix \eqref{Smat} fulfils all of them.
Checking these properties, it will be important that the angles $\psi_{\ba}$ associated
to antiparticles differ from the particle angles as $\psi_{\ba}=\psi_a\pm \pi$.
This relation is clear from Fig. \ref{figcover}b.

The conditions imposed on a two-particle $S$-matrix include
(see, for example, \cite{Korff:2000xz,CastroAlvaredo:2003it}):
\begin{itemize}
\item {\it Lorentz invariance} --- It simply means that the $S$-matrix depends on the rapidity difference
of two particles $\theta=\theta_a-\theta_b$, what is clearly true in our case.
\item {\it Zamolodchikov algebra} --- It means that the particle creation operators must satisfy
$\Phi_a(\theta)\Phi_b(\theta')=S_{ab}(\theta-\theta')\Phi_{b}(\theta')\Phi_a(\theta)$. Applying this identity twice,
one gets the following restriction on the $S$-matrix
\be
S_{ab}(\theta)S_{ba}(-\theta)=1.
\label{ZamalgS}
\ee
In the case of unitary theories, this relation can  be seen as a combination of two conditions: {\it unitarity} and
{\it Hermitian analyticity} \cite{CastroAlvaredo:2003it}.
As easy to see, our $S$-matrix satisfies only the combination \eqref{ZamalgS}
as soon as the integration constants are chosen so that
\be
C_{ab}C_{ba}=(-1)^{\<\gamma_a,\gamma_b\>}.
\label{firstC}
\ee

\item {\it Crossing symmetry} --- It relates the scattering in the $s$- and $t$-channels and requires that
\be
S_{b\ba}(\pi \I-\theta)= S_{ab}(\theta).
\ee
Again the matrix \eqref{Smat} fulfils this constraint if
\be
C_{b\ba}=(-1)^{\<\gamma_a,\gamma_b\>}C_{ab}.
\label{crossC}
\ee

\item {\it Yang-Baxter equation} ---
It means that the order in which the particles are scattered does not matter and can be depicted as
follows
\\
\unitlength .3mm 
\linethickness{0.4pt}
\ifx\plotpoint\undefined\newsavebox{\plotpoint}\fi 
\begin{picture}(310,152)(-100,30)
\thinlines
\put(20,145){\vector(1,-1){120}}
\put(190,145){\vector(1,-1){120}}
\put(139.5,144.75){\vector(-1,-1){120}}
\put(309.5,144.75){\vector(-1,-1){120}}
\put(47.5,145){\vector(0,-1){120}}
\put(282.5,145){\vector(0,-1){120}}
\put(18.75,151){\makebox(0,0)[cc]{$a$}}
\put(188.75,151){\makebox(0,0)[cc]{$a$}}
\put(47,152){\makebox(0,0)[cc]{$b$}}
\put(282,152){\makebox(0,0)[cc]{$b$}}
\put(139,151){\makebox(0,0)[cc]{$c$}}
\put(309,151){\makebox(0,0)[cc]{$c$}}
\thinlines
\qbezier(47.5,111.25)(50.75,110.375)(52,113)
\qbezier(282,111.25)(278.75,110.375)(277.5,113)
\qbezier(47.5,45.25)(42.875,42.875)(42.75,48)
\qbezier(282.5,45.25)(287.125,42.875)(287.25,48)
\qbezier(75.75,80.75)(80,76.25)(84.25,80.75)
\qbezier(245.75,80.75)(250,76.25)(254.25,80.75)
\put(53.5,100.75){\makebox(0,0)[cc]{$\theta$}}
\put(277.5,100.75){\makebox(0,0)[cc]{$\theta'$}}
\put(43.5,37.5){\makebox(0,0)[cc]{$\theta'$}}
\put(288.5,37.5){\makebox(0,0)[cc]{$\theta$}}
\put(80,63.75){\makebox(0,0)[cc]{$\theta+\theta'$}}
\put(250,63.75){\makebox(0,0)[cc]{$\theta+\theta'$}}
\put(164.75,84.75){\makebox(0,0)[cc]{$=$}}
\end{picture}
\be
S_{a b}(\theta)S_{a c}(\theta+\theta')S_{b c}(\theta')
=S_{b c}(\theta')S_{a c}(\theta+\theta')S_{a b}(\theta),
\label{YBeq}
\ee

When particles do not have additional degrees of freedom, as in our case, the $S$ matrix is
 purely diagonal and  this equations is trivially satisfied.

\item {\it Bootstrap identity} --- This is the most non-trivial requirement on the $S$-matrix which relates
its singularity structure to the spectrum. Namely, it demands that if $S_{ab}(\theta)$ has
a pole\footnote{Usually it is taken in the strong form of a pole of order one,
but experience in the field of $S$ matrix shows that this principle has to be usually extended to poles
of order greater than one \cite{Dorey:1996gd}.}
in the physical strip, {\it i.e.}, at $\theta=\I u_{ab}^{c}$ where $u_{ab}^c\in(0,\pi)$, then
the spectrum should contain the bound state $\bc$
with the mass
\be
m_{\bc}^2=m_a^2+m_b^2+2m_a m_b \cos u_{ab}^c,
\label{massc}
\ee
appearing in the fusing process $a+b\to \bc$.
The crossing symmetry then leads to the existence of other two fusing processes,
$b+c\to\ba$ and $c+a\to \bb$, with the fusing angles satisfying
\be
u_{ab}^c+u_{bc}^a+u_{ca}^b=2\pi.
\label{fussum}
\ee
But the most important condition is that it does not matter whether an additional particle, say $d$,
scatters with the bound state $\bc$ or consequently with the two particles $a,b$
\\
\unitlength .3mm 
\linethickness{0.4pt}
\ifx\plotpoint\undefined\newsavebox{\plotpoint}\fi 
\begin{picture}(310,152)(-100,30)
\thinlines
\put(70,145){\line(2,-3){30}}
\put(200,145){\line(2,-3){30}}
\put(130,145){\line(-2,-3){30}}
\put(260,145){\line(-2,-3){30}}
\put(100,80){\vector(2,-3){30}}
\put(230,80){\vector(2,-3){30}}
\put(100,80){\vector(-2,-3){30}}
\put(230,80){\vector(-2,-3){30}}
\put(50,145){\vector(2,-1){100}}
\put(180,115){\vector(2,-1){100}}
\put(100,100){\line(0,-1){20}}
\put(230,100){\line(0,-1){20}}
\put(68.75,151){\makebox(0,0)[cc]{$a$}}
\put(198.75,151){\makebox(0,0)[cc]{$a$}}
\put(47,152){\makebox(0,0)[cc]{$d$}}
\put(180,122){\makebox(0,0)[cc]{$d$}}
\put(260,152){\makebox(0,0)[cc]{$b$}}
\put(130,152){\makebox(0,0)[cc]{$b$}}
\put(108,94){\makebox(0,0)[cc]{$\bc$}}
\put(238,94){\makebox(0,0)[cc]{$\bc$}}
\thinlines
\put(164.75,84.75){\makebox(0,0)[cc]{$=$}}
\end{picture}
\be
S_{da}(\theta-\I \bu_{ca}^b)S_{db}(\theta+\I \bu_{bc}^a)=S_{d\bc}(\theta),
\qquad
\bu_{ab}^c=\pi - u_{ab}^c.
\label{bootstr}
\ee

In our case the poles of the $S$-matrix \eqref{Smat} correspond to
\be
u_{ab}^c= \psi_b-\psi_a.
\label{polesS}
\ee
The condition that the pole is on the physical strip is equivalent to $\sin u_{ab}^c >0$.
The mass formula \eqref{massc} together with its expression \eqref{ident} in terms of the central charge
$Z_\gamma$ yields
\be
\beta m_{\bc}=4\pi \cR|Z_{\gamma_a+\gamma_b}|,
\ee
so that we deduce that the bound state $\bc$ has the charge $\gamma_a+\gamma_b$ consistently with the physical
interpretation.
Moreover, since $\<\gamma_b,\gamma_{c}\>=\<\gamma_c,\gamma_a\>=\<\gamma_a,\gamma_b\>$, the $S$-matrix
elements $S_{bc}$ and $S_{ca}$ have poles of the same degree as in $S_{ab}$.
By simple manipulations one finds that the corresponding fusing angles\footnote{These definitions are valid for $\psi_a<\psi_c$.
Otherwise the shift $2\pi$ will appear in one of the other fusing angles.}
$u_{bc}^a=\psi_c-\psi_b$, $u_{ca}^b=\psi_a-\psi_c+2\pi$ also
belong to the physical strip due to
\be
\sin u_{bc}^a=\frac{m_a}{m_b}\sin u_{ca}^b=\frac{m_a}{m_c}\sin u_{ab}^c>0
\ee
and satisfy the constraint \eqref{fussum}.
This means that the other two fusing processes obtained by crossing also exist.
Finally, it is straightforward to check that the bootstrap identity \eqref{bootstr}
does hold provided
\be
C_{da}C_{db}=C_{d\bc}.
\label{bootC}
\ee

The three conditions \eqref{firstC}, \eqref{crossC} and \eqref{bootC} on the integration constants fix them to be
$C_{ab}=\sigma_{ab}C^{\<\gamma_a,\gamma_b\>}$ where $\sigma_{ab}$ is a $\IZ_2$-valued function
on the square of the charge lattice satisfying the same conditions \eqref{firstC}, \eqref{crossC} and \eqref{bootC}.
Given a polarization into electric and magnetic charges, it can be chosen, for example, as
$\sigma_{ab}=(-1)^{q_{a,\Lambda} p_b^\Lambda}$. It reminds a lot the quadratic refinement
introduced in \cite{Gaiotto:2008cd} and mentioned in footnote \ref{foot_quadrref}.

\end{itemize}

Thus, our $S$-matrix \eqref{Smat} satisfies all necessary requirements of an integrable model in $1+1$ dimensions.

\subsection{Y-system}

Y-system \cite{Zamolodchikov:1991et, Ravanini:1992fi}  is a system of functional algebraic relations on the exponentials of
the spectral densities which, although equivalent to the integral equations of TBA, play an important role
in the analysis of integrable models. In particular, it may be viewed as an intermediate step
between TBA and the transfer matrix approach. Therefore, it would be nice to have such relations at our disposal.

Let us introduce the Y-functions as\footnote{Note the additional minus sign in the definition of the Y-function.
It is introduced to ensure that $Y_{n\gamma}=Y_\gamma^n$ and hints that the bosonic version of TBA might be
more relevant in this context (see the end of section \ref{sec_disc}).}
\be
Y_a(\theta)=-e^{\beta\mu_a-\epsilon_a(\theta)}.
\ee
Using $\psi_{\ba}=\psi_a+\pi$ and $\Theta_{\ba}=-\Theta_a$, it is easy to check that \eqref{eqYigi} leads to the following relations
\be
Y_a \(\theta+\frac{\pi \I}{2}\)Y_{\ba} \(\theta-\frac{\pi \I}{2}\)
=\prod_{\ba<b<a}\Bigl[1-Y_b\bigl(\theta+\I\({\textstyle{\pi\over 2}}+\psi_a-\psi_b\)\bigr)\Bigr]^{\hng{b}\<\gamma_a,\gamma_b\>}.
\label{Ysys}
\ee
This is our Y-system. Of course, for a general configuration of charges, it is extremely complicated.
We would like to point out however a few unusual features of this Y-system \eqref{Ysys} comparing
to the standard Y-systems appearing in the literature on integrable models:
\begin{itemize}
\item First, on the l.h.s. of \eqref{Ysys} one multiplies functions associated with a particle
and its antiparticle, whereas usually one has only one function.
This feature is related to the absence of the parity symmetry in our case and to the unusual reality
 conditions.
Notice that the reality conditions \eqref{realXig} in terms of the spectral densities
and the Y-functions read
\be
\bar\epsilon_a(\theta)=\epsilon_{\ba}(-\theta),
\qquad
\bY_a(\theta)=Y_{\ba}(-\theta).
\label{realY}
\ee
As a result, the Y-functions are not necessarily real on the real axis of the spectral parameter.
Neither is the combination on the l.h.s. of \eqref{Ysys}.

\item
Second, on the r.h.s. the Y-functions are all evaluated at different points, whereas usually their arguments
do not contain any shifts. Moreover, usually the fusing angles and the only shifts appearing in functional relations
are rational multiples of $\pi$. Here they are completely arbitrary and vary continuously with
the moduli $z^\Lambda$.\footnote{It is interesting that there is a case where many of these complications
disappear. We present it in appendix \ref{sec_example}. This is probably one of the simplest possible
examples of the TBA systems of the type considered in this paper.}

\item Third, usually the power of each element of the product on the r.h.s is related to the incidence matrix of a
graph which structure is severely constrained by the periodicity of the Y-system \cite{Ravanini:1992fi,Frenkel:1995vx}.
This incidence matrix is equivalently described by the matrix
$N_{ab}=\int^{+\infty}_{-\infty} \phi_{ab}(\theta)d\theta$. In our case it is just not defined
because the kernel is not integrable.
However, this describes only a particular class of the Y-systems which are known nowadays.
It is known that the general mathematical structure
which lies behind Y-systems is related to cluster
algebras \cite{FominZelevinsky,Kuniba:2009rh}.
Such general Y-systems are defined by skew-symmetrizable matrices
having to satisfy some mutation identities. Our matrix $\hng{b}\langle \gamma_a, \gamma_b\rangle$ is also skew-symmetrizable
and originates in some mathematical structures also having relations with cluster transformations \cite{ks}.
Therefore, it would be quite interesting to understand what is
the precise connection between cluster algebras and the Y-system appearing in our case.

\end{itemize}

\subsection{Remarks on the strong coupling limit}
\label{subsec_limit}

Usually, TBA is a very effective tool to get the conformal or high temperature limit of the
integrable model.
This is a limit where the parameter $\beta$ \eqref{Bethe} goes to zero.
Given the identification \eqref{ident}, this is equivalent to vanishing of the parameter $\cR$,
which means either the small radius limit in gauge theory, where it becomes effectively three-dimensional,
or the strong coupling limit in string theory, where the ten-dimensional string coupling becomes large.

The standard derivation   tells us that in this limit the Y-functions become constant and real for a wide range of
the rapidity parameter $\theta$   \cite{Zamolodchikov:1989cf}. The values of these
functions can then be easily found from
the Y-system which reduces to a system of algebraic equations.
Finally, there is a nice formula
\be
\cF(\beta)= -\frac{1}{\pi}\sum_a \cL(Y_a) + O(\beta)
\label{limF}
\ee
which gives the free energy in terms of these constant values, where $\cL$ is the Rogers dilogarithm
defined by $\cL(x)={\rm Li}_2(x)+\frac{1}{2} \log(x) \log(1-x).$
Thus, if the same story was valid in our case, we could hope to find the strong coupling
limit of, for example, the contact potential in an easy way.

However, the TBA \eqref{eqYigi} has several features which distinguish it from the usual integrable models
and make the story much more complicated.
In particular, one has:
\begin{itemize}
\item
more complicated reality conditions \eqref{realY} than the
ones appearing usually,
\item
the kernel $\phi_{ab}(\theta)$ \eqref{ident} is not decaying at infinity,
\item
the system is supplied by arbitrary imaginary chemical potentials $\mu_a$.
\end{itemize}
On the other hand, the standard derivation
of the free energy in the conformal limit relies on the absence of these features.
In fact, it can be extended to include some of them. For example, \cite{Fendley:1991xn}
gave a generalization for non-vanishing chemical potentials leading to
the same formula \eqref{limF} with a simple modification of the Rodgers function
${\cal L}$: in $\log Y_a$ in the second term one should subtract the contribution of the chemical
potential, {\it i.e.}, it can be replaced by $-\eps_a$.
But altogether the above features give rise to the appearance of new phenomena.
Most importantly is that the Y-functions are not constant anymore in this limit.
This can be seen analytically and has been also verified by a numerical analysis
of the simplest example proposed in appendix \ref{sec_example}.
As a result, the derivation of the free energy must be seriously reconsidered and
we leave the detailed investigation of this issue for a future work.

There is also another problem to handle in order to find the  strong coupling limit of the free energy, which
 comes from the fact that the lattice of charges is infinite.
Indeed, the formula \eqref{limF} gives the free energy as a sum over all particles in the spectrum
so that we will have to sum over all charges. This sum is usually divergent
and requires a certain resummation \cite{Pioline:2009ia}. But the resummation should be performed
{\it before} the limit
since these two procedures are not commuting.

This issue can easily be exemplified if one considers the contributions of only D(-1)-instantons
to the hypermultiplet moduli space of string theory. In this case TBA can be solved exactly
since the scalar product of two charges is always vanishing leaving us with
\be
\eps_{q}(\theta)=4\pi \cR|q|\cosh\theta.
\ee
Then plugging this result into \eqref{grstener} and extracting the limit
or using \eqref{limF} with $Y_q=e^{-2\pi\I q \zeta^0}$, one finds (with $n_q=\chi_X$)
\be
\cF(0)= -\frac{\chi_X}{\pi}\sum_{q\ne 0}  {\rm Li}_2\(e^{-2\pi\I q \zeta^0}\)
\label{limFq}
\ee
which is clearly divergent. On the other hand, the same limit can be found by a Poisson resummation
of the initial expression obtained by taking the integral in \eqref{grstener}
explicitly in terms of Bessel functions \cite{RoblesLlana:2006is,Alexandrov:2008gh}.
This leads to \cite{Alexandrov:2009qq}
\be\label{strong}
\cF(\beta) = -\frac{\chi_X}{4\pi}\,   \zeta(3) \beta^{-1}
+ O(\beta),
\ee
where we identified $\beta$ with $\cR$. The leading contribution has a different scaling in $\beta$  than
the one of the usual result \eqref{limF} and reflects the divergence of the latter.

Thus, the study of the strong coupling limit is supplied here with two problems:
first, our TBA is much more general than one usually considers and, second, the infinite spectrum
requires a resummation technique to be applied.

\section{Discussion}
\label{sec_disc}

In this paper we demonstrated that the relation noticed in \cite{Gaiotto:2008cd} between
the equations describing the non-perturbative geometry of moduli spaces and
the equations of TBA goes beyond the formal analogy.
To this end, we provided an identification of the physically relevant quantities at the two sides
of the correspondence, showed that the $S$-matrix underlying this TBA fulfills all usual constraints
imposed by integrability, and derived the associated Y-system.

Of course, these results are only a first small step towards a deeper understanding of this relation.
However, already at this point we seem to open new interesting connections with other developments.
First, quite similar TBA equations and Y-systems to those considered in this paper
appear in the context of minimal area surfaces in $AdS_5$ \cite{Alday:2010vh,Hatsuda:2010cc}.
In fact, the TBA for the full $AdS_5$ problem found in \cite{Alday:2010vh} possesses all features
listed in section \ref{subsec_limit} which complicate the evaluation of the conformal limit.
Nevertheless, it has been successfully computed in \cite{Alday:2010vh}.
This was possible due to an additional $\IZ_4$ symmetry, which considerably simplifies the corresponding
Y-systems, and additional restrictions on the chemical potentials. The TBA with generic chemical potential
seems to experience the same phenomena which were mentioned in section \ref{subsec_limit}
and thus requires much more care.

Second, the Yang--Yang functional, providing for us the K\"ahler potential
on the gauge theory moduli space, plays also a prominent role in the relation of BPS vacua of certain $N=2$ gauge theories
to Bethe states of integrable models \cite{Nekrasov:2009rc}.
In that correspondence it has been identified with the twisted effective superpotential
of the low energy effective theory. Since the context of the two stories is very similar,
it is tempting to assume that the Kahler potential and the twisted effective superpotential are also related.

Besides, a somewhat intriguing observation is that the Y-system presented in appendix \ref{sec_example}
leads to the MacMahon function, the generating function of 3d partitions.
This hints that there might be a relation with the beautiful duality between
melting crystals and topological strings \cite{Okounkov:2003sp}.

A set of interesting questions arises if one tries to draw physical consequences from the fact that the $S$-matrix \eqref{Smat}
is consistent with the bootstrap. Usually, the latter can be used to generate the complete  spectrum out of some
``elementary" particles. Can it be used in the same way, for example, for gauge theory?
Which singularities of the $S$-matrix should be taken into account, all or only those which
belong to the physical strip, as in the main text? Do zeros of the $S$-matrix play some role?

The last two questions become especially relevant, if one remarks that there are actually two versions of TBA,
for fermionic and bosonic particles. Here we used the fermionic version. The bosonic equations differ
only by two signs: one should flip the sign in front of the integral in \eqref{Bethe} and
the sign in the logarithm inside the integral. This does not change much. The second sign leads
to disappearance of the shift by $\pi \I$ in the chemical potential $\mu_a$ \eqref{ident}.
But the first sign gives the $S$-matrix which is the inverse of \eqref{Smat}. As a result, the zeros
and poles are exchanged between each other. What version of TBA is relevant for our problem?
Is it related to the signs of the invariants $n_\gamma$ or to the quadratic refinement of \cite{Gaiotto:2008cd}?
Is it important at all? These are just few questions which arise naturally.

\acknowledgments

We are grateful to N. Cramp\'e, V. Fateev, N. Gromov, V. Kazakov, N. Kitanin, I. Kostov, B. Pioline, A. Sever, S. Vandoren,
P. Vieira, K. Zarembo and M. Zvonarev for valuable discussions.
The research of S.A. was supported in part by Perimeter Institute for Theoretical Physics where this work was finished.
S.A. thanks Perimeter Institute for the kind hospitality and the financial support.

\appendix

\section{The K\"ahler potential}
\label{ap_KP}

In this appendix we derive the exact instanton corrected K\"ahler potential \eqref{Kahlerinst}
and after that discuss its transformation under symplectic symmetry.

\subsection{Evaluation}

The starting point for evaluation of the K\"ahler potential is
the representation \eqref{Kdef}, which in our case takes the following form
\bea
K_\cM &=&
-\frac{1}{4\pi}\oint_{\zeta=0} \!\!\frac{\de \zeta}{\zeta^3}\(F(\zeta\etai{+})-\zeta A^\Lambda F_\Lambda(\zeta\etai{+})\)
-\frac{1}{4\pi}\oint_{\zeta=\infty} \!\!{\de \zeta}\,\zeta\(\bF(\zeta^{-1}\etai{-})
-\zeta^{-1} A^\Lambda \bF_\Lambda(\zeta^{-1}\etai{-})\)
\nn\\
&& +\frac{1}{16\pi^3}\sum_a \hng{a} \int_{\ellg{\gamma_a}}\!\! \frac{\d \zeta}{\zeta}
\[\Li_2 \(e^{-2\pi \I \Xigi{a}}\)+\frac{\hng{a}}{2}\,q_{a,\Lambda}p_a^\Lambda\(\log\(1-e^{-2\pi \I \Xigi{a}}\)\)^2
\right. \nn \\
&& \left.
+2\pi \I p_a^\Lambda \(\mui{a}_\Lambda-B_\Lambda\)\log\(1-e^{-2\pi \I \Xigi{a}}\)\],
\label{reprKP}
\eea
where $\Xigi{a}=q_{a,\Lambda}\etai{a}^\Lambda-p_a^\Lambda\mui{a}_\Lambda$, the twistor lines $\etai{a}^\Lambda,\mui{a}_\Lambda$
can be read off from \eqref{exline}, \eqref{sftwistl}, and the coordinates $A^\Lambda,B_\Lambda$ are related
to the physical fields as follows \cite{Alexandrov:2009zh}
\be
A^\Lambda=\zeta^\Lambda,
\qquad
B_{\Lambda}=\tzeta_\Lambda -\zeta^\Sigma \Re F_{\Lambda\Sigma}
-\frac{\I\Im F_{\Lambda\Sigma}}{8\pi^2}\sum_a \hng{a} p_a^\Sigma \Igg{a}(0).
\ee
The expressions for $\etai{\pm}^\Lambda$ will not be needed explicitly. For our purposes it is enough to know that
\be
\etai{\pm}^\Lambda(\zeta)=\etai{a}^\Lambda(\zeta)+O(\zeta^{\pm 2}).
\label{relXXexp}
\ee
Plugging all definitions into \eqref{reprKP}, evaluating the first two integrals by residues
and taking into account the quasi-homogeneity property \eqref{quasiF}, it is straightforward to obtain
that the K\"ahler potential is given by the first two lines in \eqref{Kahlerinst} plus the following contribution
\be
-\frac{1}{32\pi^3}\sum_a \hng{a}p_a^\Lambda \int_{\ellg{\gamma_a}}\!\! \cD_a\zeta
\[\hng{a} q_{a,\Lambda} \log\(1-e^{-2\pi \I \Xigi{a}}\)
-\frac{1}{2\pi\I}\sum_b \hng{b}q_{b,\Lambda} \int_{\ellg{\gamma_b}}\!\! \cD_b\zeta'\, \frac{\zeta+\zeta'}{\zeta-\zeta'}\]
\label{addcontr}
\ee
with the measure $\cD_a\zeta$ defined in \eqref{measure}. First, let us concentrate on the second term for $b\ne a$.
The double integral is antisymmetric in $a,b$ so that it reproduces the remaining contribution to
$K_\cM$ from the third line of \eqref{Kahlerinst}. Second, the term with $b=a$ seems to be singular.
However, it is easy to realize that it can be represented as a difference of two double integrals such that
the first integral goes over a contour which is either to the left or to the right of the second.
As a result, it gives just the residue at $\zeta'=\zeta$ which exactly cancels the first term in \eqref{addcontr}.

\subsection{Symplectic invariance}

The K\"ahler potential on the moduli space is expected to respect the symplectic invariance of the gauge theory.
However, the potential \eqref{Kahlerinst} is clearly not invariant under symplectic transformations.
This is true already at the tree level. This phenomenon was explained in \cite{DeJaegher:1997ka}
where it was shown that $K_\cM$ is actually invariant up to a K\"ahler transformation so that the moduli space
metric does not change.
Here we would like to generalize this result to the non-perturbative level.

To this end, let us review how the symplectic transformations affect various quantities \cite{deWit:1992wf,DeJaegher:1997ka}.
These transformations are represented by $2d\times2d$ matrices acting on the symplectic vectors as
\be
\label{emag}
\begin{pmatrix}
z^\Lambda\\
F_\Lambda
\end{pmatrix} \mapsto
\begin{pmatrix} \CA & \CB \\ \CC & \CD \end{pmatrix}
\begin{pmatrix}
z^\Lambda\\
F_\Lambda
\end{pmatrix}
\ee
whose blocks satisfy
\be
\begin{array}{c}
\CA^{\rm T} \CB - \CB^{\rm T} \CA =
\CA^{\rm T} \CC - \CC^{\rm T} \CA =
\CB^{\rm T} \CC - \CC^{\rm T} \CB =
\CB^{\rm T} \CD - \CD^{\rm T} \CB = 0 ,
\\ \vphantom{\mathop{A}\limits^{A^A}}
\CA^{\rm T} \CD - \CC^{\rm T} \CB =  \bf{1} .
\end{array}
\ee
It is convenient to introduce the holomorphic matrices
\be
{\cS^\Lambda}_\Sigma(z)={\CA^\Lambda}_\Sigma+\CB^{\Lambda\Theta}F_{\Theta\Sigma}(z),
\qquad
\cZ^{\Lambda\Sigma}(z)={[\cS^{-1}(z)]^\Lambda}_\Theta \CB^{\Theta\Sigma}.
\ee
The nice feature of the matrix $\cZ$ is that it is symmetric whereas $\cS$
encodes transformation properties of various quantities such as
\be
\begin{split}
F_{\Lambda\Sigma}\ \mapsto\ & ({\CD_\Lambda}^\Xi F_{\Xi\Theta}+\CC_{\Lambda\Theta}){[\cS^{-1}]^\Theta}_\Sigma,
\\
\Im F_{\Lambda\Sigma}\ \mapsto\ & \Im F_{\Theta\Xi}{[\cS^{-1}]^\Theta}_\Lambda {[\bar\cS^{-1}]^\Xi}_\Sigma,
\\
\tzeta_\Lambda-F_{\Lambda\Sigma}\zeta^\Sigma\ \mapsto\ & \(\tzeta_\Theta-F_{\Theta\Sigma}\zeta^\Sigma\){[\cS^{-1}]^\Theta}_\Sigma,
\\
q_\Lambda-F_{\Lambda\Sigma}p^\Sigma\ \mapsto\ & \(q_\Theta-F_{\Theta\Sigma}p^\Sigma\){[\cS^{-1}]^\Theta}_\Sigma.
\end{split}
\label{symptrans}
\ee
Using these results, one can show that the variation of the K\"ahler potential \eqref{Kahlerinst} under
a finite symplectic transformation is given by
\be
\Delta K_\cM=
-\frac{\I}{4}\, \cZ^{\Lambda\Sigma}(z) w_\Lambda w_\Sigma+\frac{\I}{4}\, \bar\cZ^{\Lambda\Sigma}(\bz) \bw_\Lambda \bw_\Sigma,
\ee
where
\be
w_\Lambda\equiv \mui{+}_\Lambda|_{\zeta=0}=\tzeta_\Lambda-F_{\Lambda\Sigma}\zeta^\Sigma
-\frac{1}{8\pi^2}\sum_a \hng{a}(q_\Lambda-F_{\Lambda\Sigma}p^\Sigma)\int_{\ellg{\gamma_a}}\!\! \cD_a\zeta
\label{defholw}
\ee
together with $z^\Lambda=-\I\nui{+}^\Lambda|_{\zeta=0}$ provide the complex coordinates on the moduli space $\cM$.
This demonstrates that the change of $K_\cM$ is described by a K\"ahler transformation in agreement with symplectic
invariance.

\section{Example: rigid Calabi-Yau}
\label{sec_example}

Let us consider a particular simple case of a compactification on a rigid Calabi-Yau $X$ with
the vanishing Hodge number $h_{2,1}(X)$.
Then the hypermultiplet sector consists only from one hypermultiplet known as the universal hypermultiplet.
As a result, the lattice of charges is two-dimensional, $\gamma=(q,p)$.
In addition, the holomorphic prepotential is simply $F(X)=-\frac{\tau}{2}\, X^2$ where $\tau$ is
a fixed complex coefficient determined by the holomorphic 3-form of the Calabi-Yau \cite{Bao:2009fg}.
As a result, the central charge function becomes
\be
Z_\gamma =  q +\tau p .
\label{WUHM}
\ee

Let us restrict ourselves only to the two sets of charges: pure ``electric" $(q,0)$ and pure ``magnetic" $(0,p)$.
In other words, we ignore all possible ``dyons" with both electric and magnetic charges non-vanishing.
{}From \eqref{WUHM} it is clear that in this case there are only four sets of angles
in the game ($q,p>0$)\footnote{One could ask whether the restriction to only electric and magnetic charges
is consistent with the bootstrap, which as we know leads to the bound states of charges $\gamma_a+\gamma_b$.
Remarkably, for $\Im\tau>0$, all fusing angles corresponding to the {\it poles} of the $S$-matrix \eqref{Smat}
with $\psi_a$ from \eqref{example_fuseang} turn out to be outside of the physical strip. Therefore, if
such poles are not required to satisfy the bootstrap identities, our restriction is consistent.}
\be
\psi_{(q,0)}=0,
\qquad
\psi_{(0,p)}=\psit,
\qquad
\psi_{(-q,0)}=\pi,
\qquad
\psi_{(0,-p)}=\psit+\pi,
\label{example_fuseang}
\ee
where $\psit=\arg\tau$.
Since $Y_{q,0}=Y_{1,0}^q$ and $Y_{0,p}=Y_{0,1}^p$, in fact one has only two unknown functions.
As a result, in this sector the TBA equations read as follows
\bea
\label{eqYqp}
\begin{split}
& 4\pi \cR \cosh\theta+2\pi \I\zeta= -\log Y_{1,0}(\theta)
\\
& \quad
-\frac{\I}{4\pi}\sum_{p>0} n_{0,p} p \int\limits_{-\infty}^\infty {\d \theta'}\[
\frac{e^{\theta-\theta'}+\tap}{e^{\theta-\theta'}-\tap}\,\log\(1-Y_{0,1}^p(\theta')\)
-\frac{e^{\theta-\theta'}-\tap}{e^{\theta-\theta'}+\tap}\,\log\(1-\bY_{0,1}^p(-\theta')\)
\],
\\
&
4\pi \cR \cosh\theta-2\pi \I\tzeta= -\log Y_{0,1}(\theta)
\\
& \quad
+\frac{\I}{4\pi}\sum_{q>0} n_{q,0} q \int\limits_{-\infty}^\infty {\d \theta'}\[
\frac{e^{\theta-\theta'}+\btap}{e^{\theta-\theta'}-\btap}\,\log\(1-Y_{1,0}^q(\theta')\)
-\frac{e^{\theta-\theta'}-\btap}{e^{\theta-\theta'}+\btap}\,\log\(1-\bY_{1,0}^q(-\theta')\)
\],
\end{split}
\eea
where we defined $\tap=e^{\I\psit}$.
The corresponding Y-system \eqref{Ysys} takes in this case the following form
\be
\begin{split}
Y_{1,0} \(\theta+\frac{\pi \I}{2}\)Y_{-1,0} \(\theta-\frac{\pi \I}{2}\)
&=\prod_{p>0}\Bigl[1-Y_{0,1}^p\(\theta+\I\({\textstyle\frac{\pi}{2}}-\psit\)\)\Bigr]^{p n_{0,p}},
\\
Y_{0,1} \(\theta+\frac{\pi \I}{2}\)Y_{0,-1} \(\theta-\frac{\pi \I}{2}\)
&=\prod_{q>0}\Bigl[1-Y_{-1,0}^q\(\theta-\I\({\textstyle\frac{\pi}{2}}-\psit\)\)\Bigr]^{q n_{q,0}}.
\end{split}
\label{Ysyspq}
\ee

Especially simple the above equations become when the parameter $\tau$ is pure imaginary, {\it i.e.}, $\psit=\pi/2$.
Then, all angles \eqref{example_fuseang} are multiples of $\pi/2$ and
all fusing angles are multiples of $\pi$ and moduli independent, as in
the usual integrable models!
Moreover, in this case the Y-system loses one of its unusual features. Namely,
in \eqref{Ysyspq}, as in the standard Y-systems, all functions on the r.h.s. appear without any
shifts in the arguments.

To analyze the resulting Y-system, it is useful to take into account that $n_{q,0}=\chi_X$
\cite{Alexandrov:2008gh}. One may assume
that $n_{0,p}$ is also $p$-independent. Then the r.h.s. of \eqref{Ysyspq}
is described by the MacMahon function\footnote{We thank Boris Pioline for this remark.}
\be
S(y)=\prod_{n=1}^\infty \frac{1}{\(1-y^n\)^{n}}.
\ee
This is the generating function of plane partitions which describes also
the large volume limit of the topological string partition function as $Z(g_s)=[S(e^{-g_s})]^{\chi_X/2}$ \cite{Gopakumar:1998ii}.
This fact played the prominent role in the duality between topological strings and melting crystals \cite{Okounkov:2003sp}.
Remarkably, in our example not only this function appears,
but even the power is also given by the Euler characteristic of the Calabi-Yau.
This suggests that there might be a deep interplay between our story on one side
and random partitions and topological strings on the other.

To exemplify some properties of the TBA systems introduced in this paper, it might be useful
to consider a further truncation of this example, where one drops
the sum over the infinite set of charges and considers, for example, only charges with
$q=\pm 1$ or $p=\pm 1$. Such a truncated system is particularly suitable for numerical analysis,
but still possesses most of the non-trivial features of the full problem such as
non-decaying kernels, arbitrary chemical potentials, non-trivial reality conditions, {\it etc.}


\end{document}